\documentclass[conference]{IEEEtran}
\usepackage{xcolor,soul,framed}
\colorlet{shadecolor}{yellow}
\usepackage[pdftex]{graphicx}
\graphicspath{{../pdf/}{../jpeg/}}
\DeclareGraphicsExtensions{.pdf,.jpeg,.png}
\usepackage[export]{adjustbox}
\usepackage{multirow}
\usepackage{amsmath,amssymb,amsfonts,amsthm}
\usepackage{cite}
\usepackage{textcomp}
\usepackage{balance}
\usepackage{makecell}
\usepackage{enumitem}
\usepackage{float}
\usepackage{array}
\usepackage{mdwmath, mdwtab, eqparbox}
\usepackage{url}
\usepackage{comment}
\usepackage[caption=false,font=footnotesize]{subfig}
\usepackage{mathtools}
\usepackage{algorithm}
\usepackage{algpseudocode}
\usepackage{nomencl}

\begin{document}
    \title{Impact of Solar Integration on Grid Security: Unveiling Vulnerabilities in Load Redistribution Attacks }

\author{
    \IEEEauthorblockN{Praveen Verma\textsuperscript{1}, Di Shi\textsuperscript{1}, Yanzhu Ye\textsuperscript{2}, Fengyu Wang\textsuperscript{1}, Ying Zhang\textsuperscript{3}}
    \IEEEauthorblockA{
        \textsuperscript{1}\textit{Klipsch School of Electrical and Computer Engineering, New Mexico State University}, Las Cruces, USA \\
        \textsuperscript{2}\textit{Research and Development Division, Hitachi America Ltd}, California, USA \\
        \textsuperscript{3}\textit{School of Electrical and Computer Engineering, Oklahoma State University}, Stillwater, USA \\
        \{pverma, dshi, fywang\}@nmsu.edu, yanzhu.ye@hal.hitachi.com, y.zhang@okstate.edu
    }
\vspace{-24pt}


}
 

\maketitle
\vspace{-24pt}

\begin{abstract}
Load redistribution (LR) attacks represent a practical and sophisticated form of false data injection (FDI) attacks, where the attacker manipulates grid data to influence economic operations of the grid through misleading security constrained economic dispatch (SCED) decisions. Traditionally, LR attack models operate under the assumption that generator measurements are secure and immune to tampering. However, the increasing integration of solar generation into power grids challenges this assumption, exposing new vulnerabilities. This paper proposes an enhanced load redistribution attack model, addressing new vulnerabilities introduced by the increasing integration of solar generation in power grids. The study demonstrates that manipulating solar generation data significantly disrupts grid economics, with peak impacts during periods of high solar generation.

\end{abstract}

\vspace{0.2cm}

\makenomenclature
\mbox{}
\nomenclature[]{$\mathbf{S}\big, \mathbf{U}\big,\mathbf{V}$}{Incident matrix for shift factor, bus generator, and bus loads}
\nomenclature[]{$R^{0}\big/D$}{True value of solar generation\big/load}
\nomenclature[]{$\Delta R$\big/$\Delta D$}{Solar generation\big/load deviation vector}
\nomenclature[]{$P\big/R\big/J$}{Conventional generation\big/solar generation\big/load shedding vector}
\nomenclature[]{$S_{l}$}{$l^{th}$ row of shift factor matrix}
\nomenclature[]{$\tau\big/\alpha\big/\kappa\big/\Xi\big/\varrho$}{Load manipulation factor\big/generation manipulation factor\big/irradiance scaling factor\big/line selection factor\big/overloading factor}
\nomenclature[]{$\underline{P}$\big/$\overline{P}$}{Conventional generator's minimum\big/maximum limit vector}
\nomenclature[]{$\underline{R}$\big/$\overline{R}$}{Solar generator's minimum\big/maximum limit vector}
\nomenclature[]{$c_{g}$\big/$c_{r}$\big/$c_{d}$ }{Cost coefficient vector for conventional generation\big/solar generation\big/load shedding}
\nomenclature[]{$f_{l}^{0}$\big/$f_{l}^{max}$}{Pre-attack power flow\big/maximum power flow limit of line $l$}
\nomenclature[]{$\mathcal{L}$, $\mathcal{L}_{f}$}{Set of lines, set of vulnerable lines to induce severe overflow}
\nomenclature[]{$F$\big/$\overline{F}$}{Power flow\big/maximum power flow limit vector}
\nomenclature[]{$s^{r}$}{Rated capacity of solar generation}
\nomenclature[]{$I\big/I_{s}$}{Solar irradiance\big/standard solar irradiance}
\nomenclature[]{$S_{c}$}{Irradiance point}


\section{Introduction}\label{sec:intro}
\IEEEPARstart{L}{oad} redistribution (LR) attacks are a sophisticated form of false data injection attacks in which false load and line power flow data are injected into the original measurements transmitted from remote locations to the control center via Information and Communication Technology (ICT) systems. By exploiting vulnerabilities in these ICT systems, attackers can manipulate the data received by grid operators, misleading the decision-making of security-constrained economic dispatch (SCED) problems and increasing the operation costs.

The concept of LR attacks was first introduced in~\cite{5754636} in 2011 as a bi-level attacker-defender optimization problem. In this framework, the attacker at the upper level aims to maximize the operation costs of the power grid by strategically manipulating measurements. In contrast, the defender at the lower level seeks to minimize the operation cost by responding to the false data. Since then, LR attack models have evolved significantly regarding their mathematical formulations, objectives, and impacts on the power grid~\cite{10521762}. These models can be broadly categorized into three types: single-level models, bi-level models, and tri-level models.


In~\cite{6805238}, the authors proposed an approach that requires only partial network information of the grid to formulate the LR attack. Similarly, in~\cite{9726789}, a bi-level model was developed, relying solely on pre-attack and post-attack power dispatch information, thereby significantly reducing dependency on hard-to-access grid parameters. However, attack models~\cite{6805238,9726789} involve high computation times to estimate the LR attack vector. To overcome this limitation, single-level attack models have been explored in the literature. In~\cite{10360224}, an iterative LR attack approach is introduced, demonstrating that LR attacks can be executed with partial system manipulation while avoiding the need for complex network information. In~\cite{9000221}, the exploitable core structure of LR attacks is analyzed, revealing that attackers can achieve their objectives without solving intricate mathematical models. In~\cite{7731227}, a single-level LP model is presented to compute the false data required for an LR attack, and its uneconomic impact on SCED is validated.

Another popular mathematical formulation is a tri-level model. In the LR attack models mentioned above~\cite{10360224,9000221,7731227}, the effect of the LR attack is validated using SCED. However, it is crucial to recognize that the SCED problem is an LP problem capable of yielding multiple feasible solutions. Among these, SCED solutions that avoid line overloading remain harmless to grid operations. Consequently, the existence of multiple solutions creates a significant obstacle for attackers in formulating a successful attack strategy. For instance, if the grid operator selects a SCED solution that prevents line overloading despite manipulated load data, the attacker’s objective is thwarted, rendering the manipulation ineffective. 
In~\cite{7271100}, authors proposed a tri-level model to overcome the challenge posed by multiple SCED solutions. This approach guarantees that, irrespective of the SCED solution selected by the grid operator, at least one transmission line in the network will be overloaded, thereby ensuring the LR attack achieves its intended disruptive impact.

On the other hand, LR attack models can be classified into two categories based on their target and impact on the grid: voltage and branch power flow manipulation. In the voltage impact category, a linear power flow-based LR attack approach for voltage security is proposed in~\cite{9916983}. Similarly, the net LR attack model in~\cite{9248915} introduces a method for hidden LR attacks on nodal voltage magnitude estimation in AC distribution networks. In~\cite{9035401}, the authors propose a method to assess the vulnerability of conservation voltage reduction under LR attack impacts in unbalanced active distribution networks. For branch power flow manipulation,\cite{8338154} introduces a screening approach for cascading failures caused by branch overflows. Similarly,\cite{8434103} presents an LR attack model that maximizes overloading in large-scale attacked branches. 

Despite differences in mathematical formulations, objectives, and grid impacts, all the aforementioned LR attack models are conceived for the assumption that generator measurements are immune to manipulation~\cite{10521762}. This assumption has remained a cornerstone of LR attack models, including the recent approaches~\cite{9726789,10360224,9000221,7731227}.
The basis for this {\em strong assumption} lies in the architecture of traditional power grids, where large, centrally located conventional generators at power plants maintain direct communication with the control center. In this setup, even if an attacker were to manipulate generator measurements, grid operators at the control center could easily verify the accuracy of these measurements with the power plant control room~\cite{5754636}, enabling the detection of any discrepancies or tampering.

However, the integration of distributed solar generation into modern power grids challenges this long-standing {\em strong assumption}. Unlike conventional generators, distributed solar generation sources are small and often lack direct communication links with the control center. This poses a vulnerability in which attackers can manipulate solar generation measurements without immediate detection. 
Efforts have been made to incorporate generation measurement manipulation into LR attack models, such as in~\cite{7286402}, but these primarily focus on conventional central generators and fail to address the unique characteristics of solar generation. Solar power output fluctuates throughout the day and across seasons, introducing operational complexities that remain underexplored in existing models. Furthermore, the analysis in ~\cite{7286402} is preliminary and lacks comprehensive insights into the broader implications of manipulating solar generation within the grid.

Therefore, in this work, we extend the LR attack model to explicitly account for the impact of distributed solar generation. The main contributions of the paper are summarized as follows:

\begin{enumerate}
    \item This paper develops a single-level LR attack model that incorporates the manipulation of distributed solar generation into existing LR attack models.
    \item Through extensive simulations and analyses conducted under varying irradiance conditions throughout the day and across seasons, we uncover critical vulnerabilities in the economic operation of the smart grid introduced due to the manipulation of distributed solar generation.
    \item The performance of the proposed single-level LR attack model is compared against other \textit{state-of-the-art} LR attack models~\cite{10360224,9000221,7731227}, revealing heightened vulnerabilities in the smart grid.
\end{enumerate}

The rest of the paper is organized as follows: Section \ref{sec:formulation} presents  the mathematical formulation of the proposed LR attack model, and the SECD optimization problem used to validate and analyze the impact of the proposed LR attack model. Section \ref{sec:case study} provides a case study on the IEEE $118$-bus system, along with a detailed comparative analysis. Lastly, Section \ref{sec:conclusion} concludes the paper with key insights and findings.

\section{The Proposed Attack Model Incorporating Solar Generation}\label{sec:formulation}

\begin{figure}[h!]
    \centering
    \includegraphics[width=0.3\textwidth]{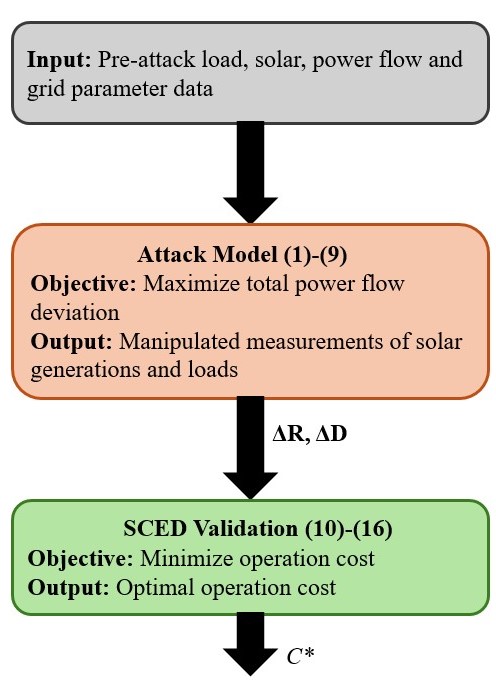}
    \caption{Flow diagram of the proposed LR attack model}
    \label{fig:flow}
\end{figure}

This section first presents the formulation of a single-level programming model designed to estimate solar generation and load deviations within the proposed LR attack framework. Next, it provides a detailed discussion of the SCED optimization problem to validate and analyze the impact of the proposed model. Fig.~\ref{fig:flow} illustrates the flow diagram of the proposed attack model.

\subsection{Proposed Attack Model}\label{sec:proposed attack model}

The proposed LR attack model assumes that the attacker has access to the following partial information: the pre-attack true load vector $\mathbf{D}$, the pre-attack true power flow of the lines $f_{l}^{0},~\forall l \in \mathcal{L}$, parameters of the network topology, including the shift factor matrix $\mathbf{S}$, bus generator incidence matrix $\mathbf{U}$, and bus load incidence matrix $\mathbf{V}$, and solar generator parameters, such as the rated capacity of solar generation $s^{r}$ and the irradiance point $S_{c}$. 

The proposed LR attack model aims to maximize power flow deviations, $\Delta f_l,~ l \in \mathcal{L}$, to induce overflows in the line set $\mathcal{L}$ by manipulating the false solar generation vector $\mathbf{\Delta R}$ and the false load vector $\mathbf{\Delta D}$. The corresponding optimization problem is formulated as follows:

\begin{gather}
    \underset{\mathbf{\Delta R, \Delta D}}{\textbf{Max}}\mathcal{O}=\underset{l \in \mathcal{L}}{\sum}\Delta f_{l}\label{eq:upperlevelobjective}\\
     \textbf{s.t.}~\Delta f_{l}=\mathbf{S_{l}\cdot U \cdot \Delta R - S_{l}\cdot V \cdot \Delta D},~~\forall l \in \mathcal{L} \label{eq:delta f defination}
    \end{gather}
 \begin{gather}
     \mathcal{L}\longleftarrow\underset{l}{arg}\Bigg(\frac{f_{l}^{0}}{\lambda_{l} \cdot f_{l}^{max}}\geq \Xi  \Bigg),~~\lambda_{l} =\begin{dcases*}
        1 & if $f_{l}^{0} \geq 0$\\
        -1 & if  $f_{l}^{0}< 0$
        \end{dcases*} \label{eq:selected lines} \\
       \Delta f_{l}\geq \varrho \cdot f_{l}^{max}-f_{l}^{0},~~ \forall l \in \mathcal{L}_{f} \& \mathcal{L}_{f}  \subset \mathcal{L} \label{eq:subset overloading}\\  
        \mathbf{1^{T}\Delta D} = 0\label{eq:demand deviation zero}\\
       \mathbf{1^{T}\Delta R} = 0\label{eq:generation deviation zero}\\
        -\tau \mathbf{D} \leq   \mathbf{\Delta D} \leq \tau \mathbf{D},~0 \leq \tau \leq 1 \label{eq:demand deviation range}\\
-\alpha \mathbf{R^{0}} \leq   \mathbf{\Delta R} \leq \alpha \mathbf{R^{0}},~0 \leq \alpha \leq 1  \label{eq:generation deviation range}\\
\mathbf{R^{0}} =\begin{dcases*}
        s^{r}\frac{\mathbf{I}^{2}}{I_{s}S_{c}} & if $ 0< \mathbf{I} < S_{c}$,\\
        s^{r} & if $\mathbf{I} \geq S_{c}$,
        \end{dcases*} 
        \label{eq:solar power output}      
\end{gather}

The constraint in \eqref{eq:delta f defination} establishes the relationship between line power flow deviations and the manipulations in the solar generation vector $\mathbf{\Delta R}$ and the load vector $\mathbf{\Delta D}$. The constraint in \eqref{eq:selected lines} aims to identify the target set of lines based upon the magnitude and direction of pre-attack power flow $f_{l}^{0}$. The constraint in \eqref{eq:subset overloading} ensures that the deviation $\Delta f_{l}$ for the selected subset $\mathcal{L}_{f}$ of lines within the target set $\mathcal{L}$ is sufficient to cause severe overloading. The constraints in \eqref{eq:demand deviation zero} and \eqref{eq:generation deviation zero} ensure that the total amount of falsely injected data for the pre-attack load and solar generation remains zero. The constraints in \eqref{eq:demand deviation range} and \eqref{eq:generation deviation range} ensure that the manipulation of load and solar generation remain within their respective pre-attack ranges. Finally, the constraints in \eqref{eq:solar power output} calculate the solar generation output $\mathbf{R^{0}}$ based on the solar irradiance $\mathbf{I}$.

In the proposed attack model \eqref{eq:upperlevelobjective}-\eqref{eq:solar power output}, constraints \eqref{eq:selected lines} and \eqref{eq:subset overloading} are specifically designed to overload vulnerable lines, thereby causing significant economic impact. The constraints \eqref{eq:delta f defination}, \eqref{eq:demand deviation zero}-\eqref{eq:solar power output} ensure that the manipulated data $\Delta f_{l},~\forall l \in \mathcal{L}$, $\mathbf{\Delta R}$, and $\mathbf{\Delta D}$ remain consistent with the physical laws governing the grid, allowing the attack to evade detection by the bad data detection mechanisms employed by grid operators at the control center~\cite{verma2024defense}.

\subsection{Validation of Impact on SCED}\label{sec:validation}

To analyze and validate the impact of manipulating the solar generation vector $\mathbf{\Delta R}$ and the load vector $\mathbf{\Delta D}$, we compare the resulting SCED operation cost against that of other {\em state-of-the-art} LR attack models. The SCED optimization problem, which incorporates the manipulated measurements of solar generation $\mathbf{\Delta R}$ and load $\mathbf{\Delta D}$, is formulated as follows:
\begin{gather}
C^{*}=\mathbf{\underset{P,R,J}{\textbf{Min}}~ c_{g}^{T}P+c_{r}^{T}R+c_{d}^{T}J}\label{eq:lowerlevelobjective}\\
\textbf{s.t.}~\mathbf{1^{T}P+1^{T}R = 1^{T} D - 1^{T} J}\label{eq:supply demand balance}\\
 \mathbf{\underline{P}\leq P \leq \overline{P}} \label{eq:gen limit}\\
 \mathbf{\underline{R}\leq R \leq \overline{R}} \label{eq:renewable gen limit}\\
\mathbf{0 \leq J \leq D+\Delta D} \label{eq:load sheddinglimit}\\
\mathbf{F=S\cdot U\cdot (P + R + \Delta R) - S\cdot V \cdot (D+\Delta D-J)}\label{eq:power flow}\\
\mathbf{-\overline{F} \leq F \leq \overline{F}} \label{eq:powerflow limit}
\end{gather}

The SCED formulation \eqref{eq:lowerlevelobjective}-\eqref{eq:powerflow limit} represents the grid operator's response to the manipulations $\mathbf{\Delta R}$ and $\mathbf{\Delta D}$. The objective function \eqref{eq:lowerlevelobjective} minimizes the operation cost by adjusting $\mathbf{P}$, $\mathbf{R}$, and $\mathbf{J}$, subject to the following constraints:
\begin{itemize}
    \item \textbf{Supply-demand balance}: Constraint \eqref{eq:supply demand balance} ensures that the total power generation matches the adjusted load.
    \item \textbf{Generation limits}: Constraints \eqref{eq:gen limit} and \eqref{eq:renewable gen limit} impose the minimum and maximum limits for conventional and renewable generation, respectively.
    \item \textbf{Load shedding}: Constraint \eqref{eq:load sheddinglimit} limits the allowable load shedding to the manipulated load $\mathbf{D + \Delta D}$.
    \item \textbf{Power flow and line limits}: Constraints \eqref{eq:power flow} and \eqref{eq:powerflow limit} represent the power flow equations and enforce line flow limits.
\end{itemize}

\section{Case Study}\label{sec:case study}
\subsection{Simulation Setup}\label{sec:sim setting}
\begin{figure}[h!]
    \centering
    \includegraphics[width=0.48\textwidth]{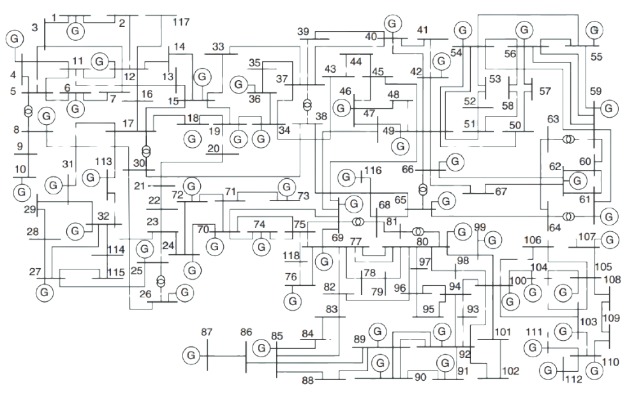}
    \caption{IEEE 118 Bus System~\cite{8939187}}
    \label{fig:test system}
\end{figure}

This case study utilizes the IEEE $118$-bus system, which comprises $99$  loads with a total demand of $4242$ MW and $54$ generators with a combined capacity of $9966$ MW. The system is shown in Fig. \ref{fig:test system}.

\begin{figure*}[t]
  \centering
    \subfloat[\small{Power output (MW) of solar generation}]{\includegraphics[width=0.495\textwidth,height=5cm,keepaspectratio]{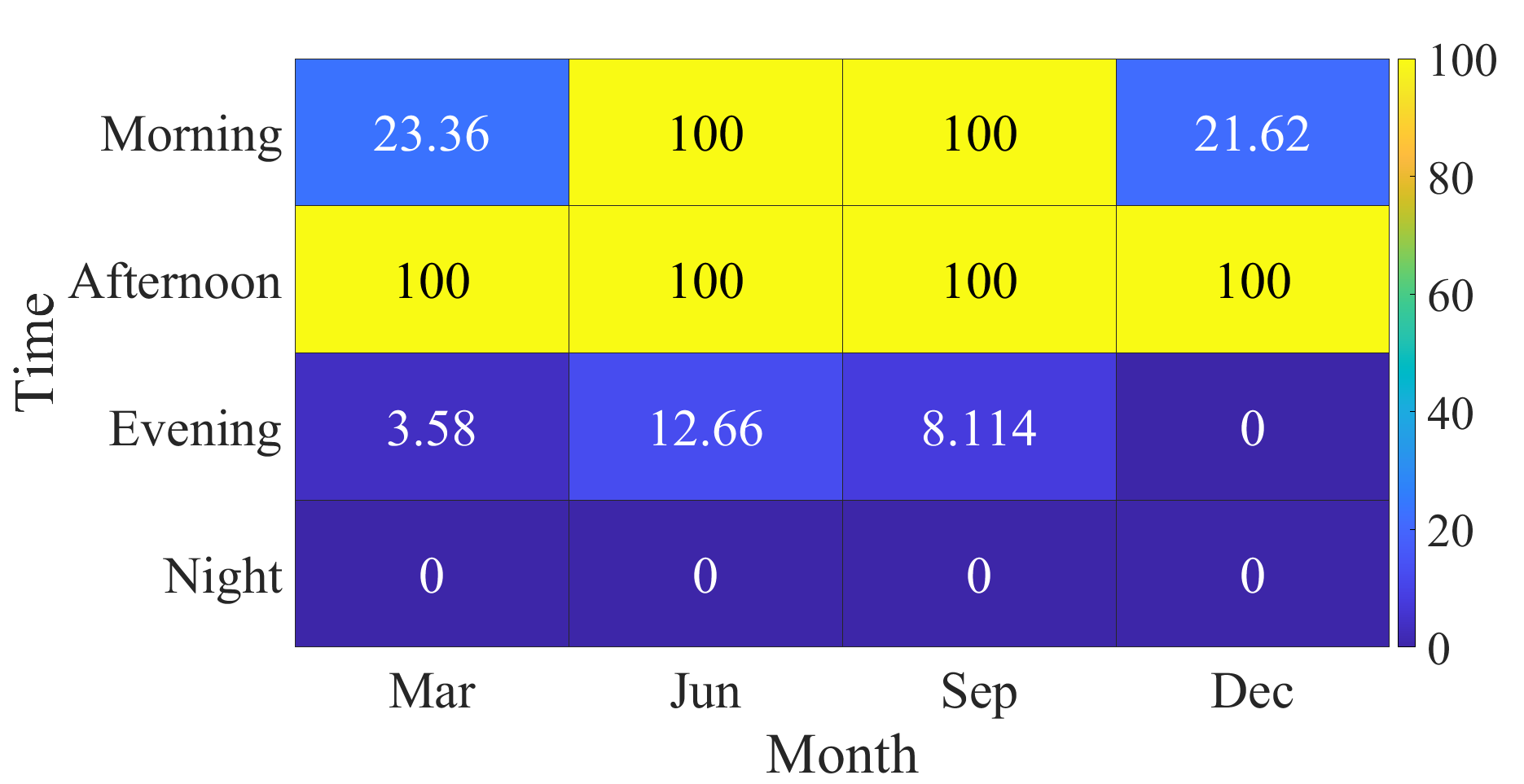}}
    \subfloat[\small{Post-attack SCED operation cost ($\$/h$)}]{\includegraphics[width=0.495\textwidth,height=5cm,keepaspectratio]{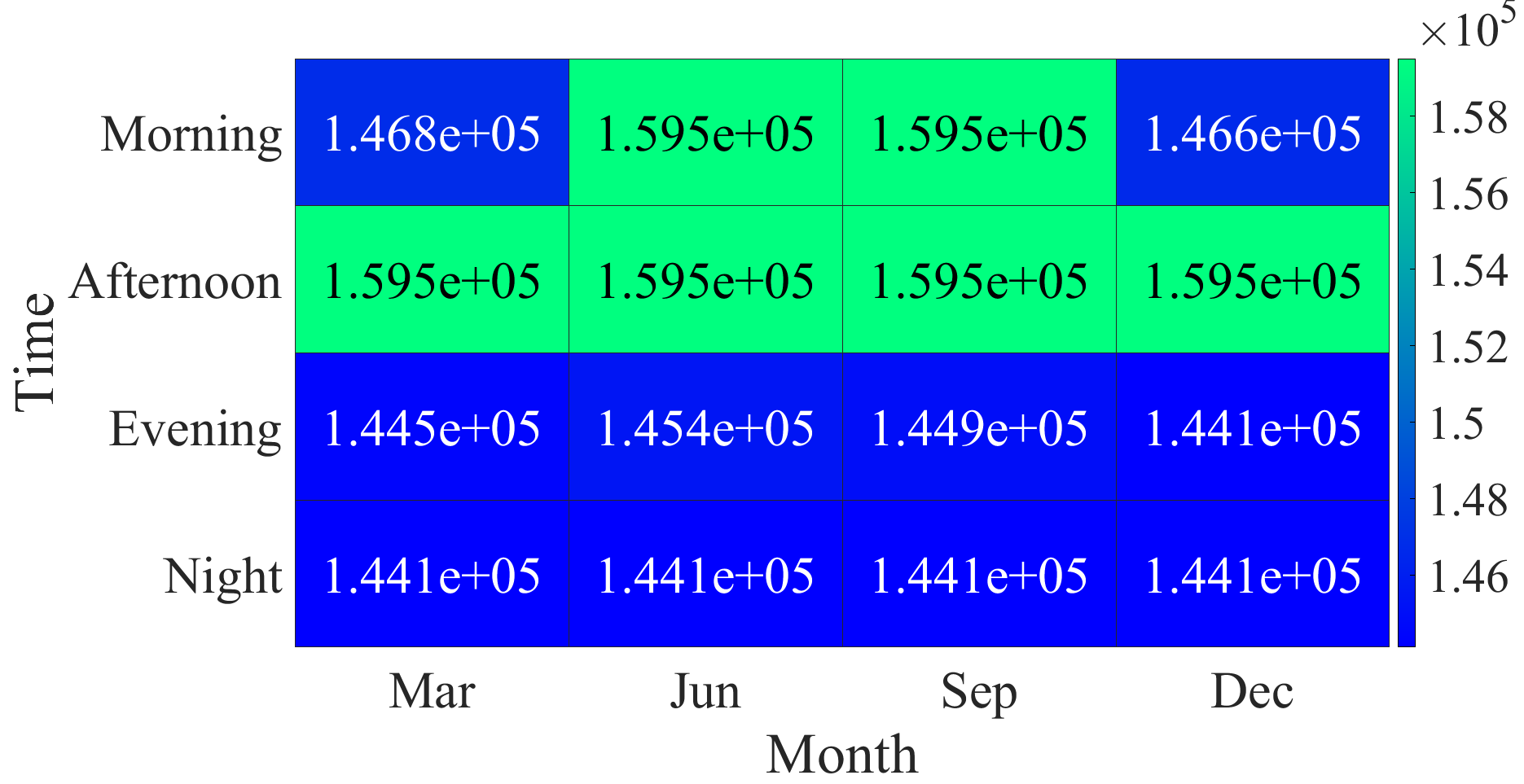}}
    \vspace{0.5cm}
    \subfloat[Increase in SCED operation cost (\$/h)]{\includegraphics[width=0.495\textwidth,height=5cm,keepaspectratio]{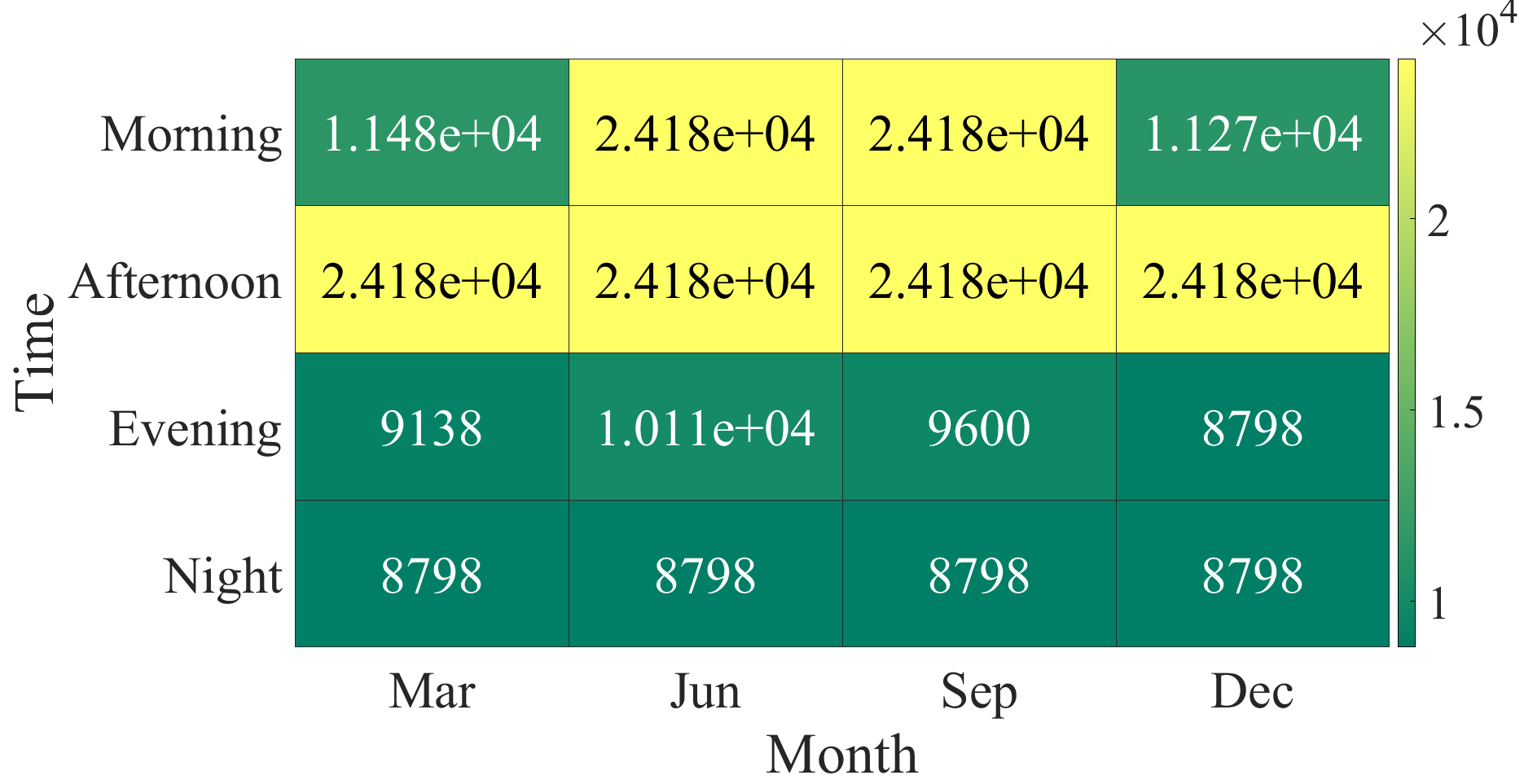}}
    \subfloat[\small{Load shedding (MW)}]{\includegraphics[width=0.495\textwidth,height=5cm,keepaspectratio]{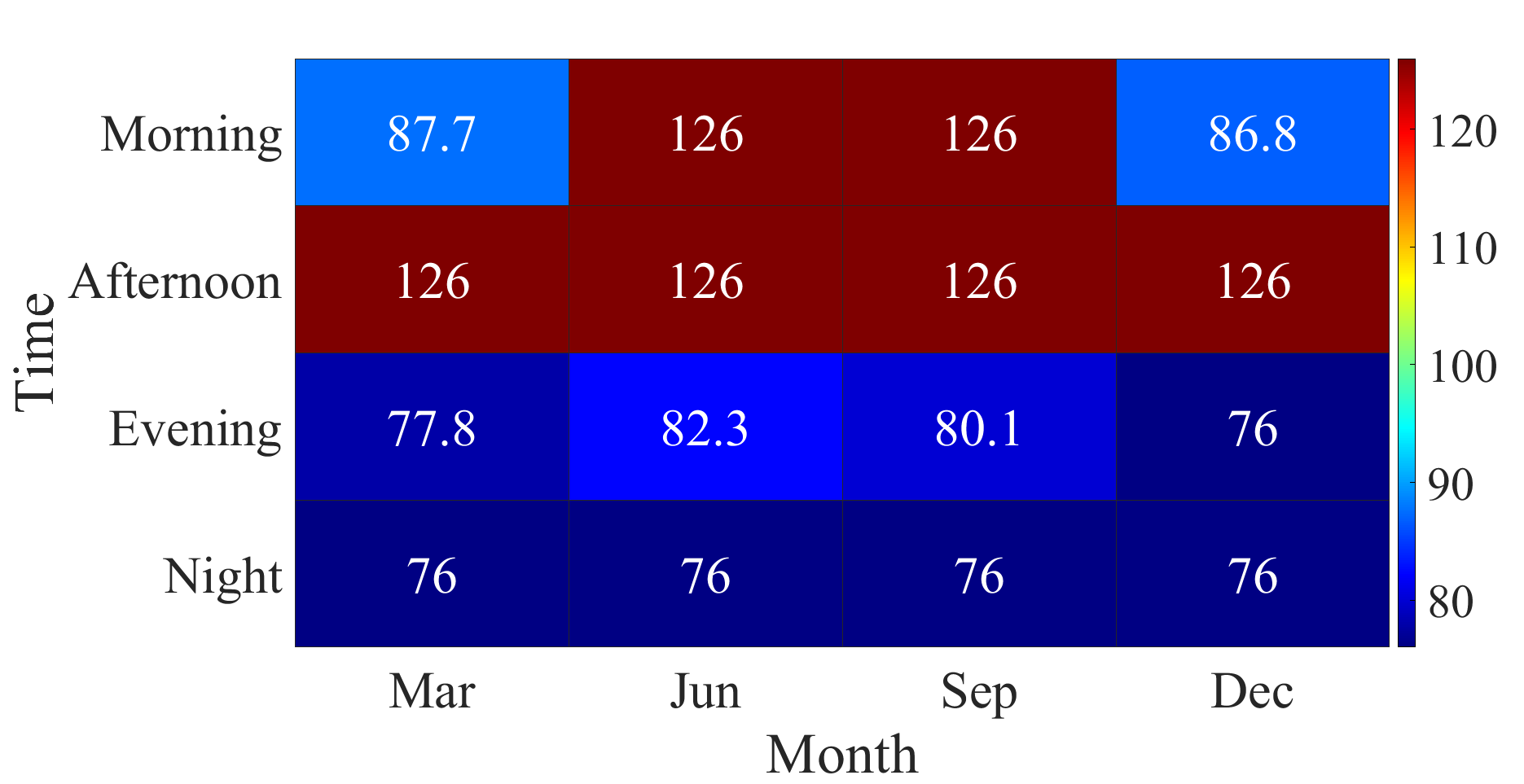}}
     \caption{{Impact of attack on solar generations on the economic operation of the grid} } \label{fig:attack impact result}
\end{figure*}

In this study, $35$ generators with a capacity of $3500$ MW are modeled as solar generators, with their rated capacity ($s^{r}$) set to match the maximum generation capacity of the non-renewable generators. The parameters $\tau$, $\alpha$, $\Xi$, and $\varrho$ are set to $0.5$, $0.5$,  $0.9$, and $1.3$, respectively. The remaining parameters are taken from the MATPOWER package~\cite{MATPOWER}.

To analyze the impact of solar generation manipulation, real-world solar data from NREL~\cite{NREL} is utilized. Four distinct days, each representing a different month and season, are selected for the study. The analysis is performed at four different times of the day: morning ($8:00$ AM), afternoon ($12:00$ PM), evening ($5:00$ PM), and night ($9:00$ PM). This results in a total of 16 scenarios, allowing for an exploration of the effects of daily and seasonal variations on the manipulation of solar generation within the proposed LR attack model.

\subsection{Impact of Manipulation of Solar Generations}\label{sec:impact}

As shown in Fig. \ref{fig:attack impact result}, daily and seasonal variations in solar power output significantly affect the post-attack SCED operation cost and load shedding.  The following observations can be made from Fig. \ref{fig:attack impact result} regarding the impact of manipulating solar generation.
\begin{enumerate}

    \item \textbf{Significant disruption to economic operations}: Manipulation of solar generation under the proposed LR attack model disrupts grid economics by increasing operation costs and triggering heightened load shedding. The magnitude of the disruption scales proportionally with the MW output of solar generation.
    
    \item \textbf{Daily variations}: The impact of manipulation is most pronounced in the afternoon, when solar generation peaks at 100 MW, as shown in Fig.~\ref{fig:attack impact result}(a). Figures~\ref{fig:attack impact result}(b), \ref{fig:attack impact result}(c), and \ref{fig:attack impact result}(d) indicate that this peak results in the worst daily impact, with a post-attack SCED operation cost of $1.595 \times 10^{5}$ \$/h, an increase of $2.418 \times 10^{4}$ \$/h compared to the pre-attack cost of $1.35 \times 10^{5}$ \$/h, and load shedding reaching $126$ MW.

    \item \textbf{Seasonal variations}: Seasonal patterns reveal heightened vulnerability during June and September, when solar power output peaks in the morning and remains high throughout the day. In these months, SCED operation costs and load shedding are particularly severe during morning and afternoon hours, with post-attack costs reaching $1.595 \times 10^{5}$ \$/h and load shedding at $126$ MW. Conversely, in March and December, the impact is less pronounced due to lower solar generation levels, especially in the morning.

    \item \textbf{Impact at night}: During nighttime across all months and in December evenings, when solar irradiance is zero and solar generation is inactive, the proposed LR attack reverts to a standard LR attack. In these scenarios, the impact is less severe compared to when solar generation is actively manipulated.

\end{enumerate}

\begin{table*}[!t]
\caption{\textsc{Comparative Study}}
\label{tab:comp study}
\centering
\begin{tabular}{|c|c|c|c|ccc|}
\hline
\multirow{3}{*}{\begin{tabular}[c]{@{}c@{}}\\ \\ Parameter\end{tabular}}     & \multirow{3}{*}{\begin{tabular}[c]{@{}c@{}}\\ \\ \cite{10360224}\end{tabular}}                                              & \multirow{3}{*}{\begin{tabular}[c]{@{}c@{}}\\ \\ \cite{9000221}\end{tabular}} & \multirow{3}{*}{\begin{tabular}[c]{@{}c@{}}\\ \\ \cite{7731227}\end{tabular}}  & \multicolumn{3}{c|}{Proposed work}                                                                                                                                                                                                                     \\ \cline{5-7} 
                                                                             &                                 &                                 &                                  & \multicolumn{1}{c|}{\multirow{2}{*}{\begin{tabular}[c]{@{}c@{}}Without solar\\  generation\end{tabular}}} & \multicolumn{2}{c|}{\begin{tabular}[c]{@{}c@{}}With solar  generation\end{tabular}}                                                      \\ \cline{6-7} 
                                                                             &                                 &                                 &                                  & \multicolumn{1}{c|}{}                                                                                     & \multicolumn{1}{c|}{\begin{tabular}[c]{@{}c@{}}Minimum   impact\end{tabular}} & \begin{tabular}[c]{@{}c@{}}Maximum  impact\end{tabular} \\ \hline
\begin{tabular}[c]{@{}c@{}}Post-attack \\ operation cost ($10^5~\$/h$)\end{tabular} &     1.37                            & 1.37                            &            1.39                     &        \multicolumn{1}{c|}{1.441}                                                                                    & \multicolumn{1}{c|}{ 1.445 }                                                        &      1.595                                                     \\ \hline
\begin{tabular}[c]{@{}c@{}}Increase in \\ operation cost ($10^3~\$/h$)\end{tabular} &       2.0                           &  2.0                          &         4.0                        &       \multicolumn{1}{c|}{8.7}                                                                                     & \multicolumn{1}{c|}{9.1}                                                          &       24.1                                                    \\ \hline
\begin{tabular}[c]{@{}c@{}}Load shedding \\ (MW)\end{tabular} &   5.28                                       &     26.16                            &         49.63                        &        \multicolumn{1}{c|}{76}                                                                                     & \multicolumn{1}{c|}{77.8}                                                          &          126                                                 \\ \hline
\end{tabular}
\end{table*}

\subsection{Comparative Study}\label{sec:comparative study}
Table~\ref{tab:comp study} compares the impact of the proposed LR attack model with the \emph{state-of-the-art} LR attack models from~\cite{10360224,9000221,7731227}. As shown in Table~\ref{tab:comp study}, even in the absence of solar generation (at nighttime), the proposed model results in a significantly higher adverse impact on the grid. Specifically, the proposed approach achieves a post-attack SCED operation cost of $1.441 \times 10^5~\$/h$, an increase of $8.7 \times 10^3~\$/h$ in operation cost, and load shedding of $76$ MW. These values are notably higher than the corresponding metrics from~\cite{10360224,9000221,7731227}. This heightened impact can be attributed to the proposed model’s ability to overload the set $\mathcal{L}$ of multiple lines and significantly overload selected lines in the subset $\mathcal{L}_{f}$ by manipulating all loads in the grid.

Incorporating solar generation manipulation into the proposed attack model further amplifies the severity of the attack. For instance, even under minimal impact scenarios, solar generation manipulation results in a post-attack SCED operation cost of $1.445 \times 10^5~\$/h$, which is substantially higher than the post-attack costs of $1.37 \times 10^5~\$/h$, $1.37 \times 10^5~\$/h$, and $1.39 \times 10^5~\$/h$ of the LR attack models of~\cite{9000221,7731227,10360224}.

Additionally, a comparative analysis of the minimal impact scenarios reveals that the proposed model leads to consistently higher increases in operation cost and load shedding compared to the LR attack models of~\cite{9000221,7731227,10360224}.

The impact is most severe when solar generation is at its peak, resulting in a post-attack SCED operation cost of $1.595 \times 10^5~\$/h$, an increase of $24.1 \times 10^3~\$/h$ in operation cost, and load shedding of $126$ MW.

\subsection{Discussion}\label{sec:discuss}
To further analyze the impact of solar generation manipulation, we consider the following scenarios:

\vspace{0.2cm}
\subsubsection{Impact of Variation in Solar Generation Manipulation Factor $\alpha$}

Table~\ref{tab:alpha impact} illustrates the impact of varying $\alpha$ on key operational metrics, such as post-attack operation costs, the increase in operation costs, and load shedding.
\begin{table}[h!]
\centering
\caption{\textsc{Impact of $\alpha$}}\label{tab:alpha impact}
\begin{tabular}{|c|c|c|c|}
\hline
$\alpha$ & \begin{tabular}[c]{@{}c@{}}Post-attack \\ operation cost\\ ($10^5~\$/h$)\end{tabular} & \begin{tabular}[c]{@{}c@{}}Increase in \\ operation  cost \\($10^3~\$/h$)\end{tabular} & \begin{tabular}[c]{@{}c@{}}Load \\ shedding \\ (MW)\end{tabular} \\ \hline
\multicolumn{1}{|c|}{0.25} & \multicolumn{1}{c|}{1.506} & \multicolumn{1}{c|}{15.3} & \multicolumn{1}{c|}{101} \\ \hline
0.50 & 1.595 & 24.1 & 126 \\ \hline
0.75 & 1.755  & 40.2 & 234.9 \\ \hline
\end{tabular}
\end{table}

For $\alpha$ = $0.25$, the post-attack operation cost rises to $1.506 \times 10^5~\$/h$, marking an increase of $15.3 \times 10^3~\$/h$ compared to the pre-attack operation cost. Additionally, the grid experiences load shedding of $101$ MW, highlighting its diminished capacity to meet demand.

As $\alpha$ increases to $0.50$, the severity of the attack escalates. The post-attack operation cost rises to $1.595 \times 10^5~\$/h$, representing a $24.1 \times 10^3~\$/h$ increase from the pre-attack cost, while load shedding increases to 126 MW, reflecting heightened system strain

At $\alpha = 0.75$, the impact becomes even more pronounced, with the post-attack operation cost surging to $1.755 \times 10^5~\$/h$, a significant increase of $40.2 \times 10^3~\$/h$, and load shedding escalating to $234.9$ MW. The numbers almost double the value observed at $\alpha = 0.50$. 

The results demonstrate a strong positive correlation between $\alpha$ and the severity of operational disruptions. As $\alpha$ increases, both economic costs and load shedding rise significantly, underscoring the grid's vulnerabilities under intensified attack conditions.

\subsubsection{Impact of Coordinated Manipulation of Solar Generation $\mathbf{R^{0}}$ with $\mathbf{\Delta D}$}
Table~\ref{tab:R0 impact} compares the impact of manipulating only the initial solar generation $\mathbf{R^{0}}$ (when $\mathbf{\Delta D} = 0$) with the coordinated manipulation of $\mathbf{R^{0}}$ and load $\mathbf{D}$ ($\mathbf{\Delta D} \neq 0$). When only $\mathbf{R^{0}}$ is manipulated (i.e., $\mathbf{\Delta D} = 0$), the adverse impact is relatively small, with a post-attack operation cost of $1.385 \times 10^5~\$/h$, an increase of $3.25 \times 10^3~\$/h$, and load shedding of $34$ MW. 

\begin{table}[h!]
\centering
\caption{\textsc{{Impact of Coordinated Manipulation}}}\label{tab:R0 impact}
\begin{tabular}{|c|c|c|c|c|}
\hline
Case & $R^{0}$ (MW) & \begin{tabular}[c]{@{}c@{}}Post-attack \\ operation cost\\ ($10^5~\$/h$)\end{tabular} & \begin{tabular}[c]{@{}c@{}}Increase in \\ operation  cost \\($10^3~\$/h$)\end{tabular} & \begin{tabular}[c]{@{}c@{}}Load \\ shedding \\ (MW)\end{tabular} \\ \hline
\multicolumn{1}{|c|}{$\mathbf{\Delta D} = 0$}&\multicolumn{1}{|c|}{100} & \multicolumn{1}{c|}{1.385} & \multicolumn{1}{c|}{3.25} & \multicolumn{1}{c|}{34} \\ \hline
$\mathbf{\Delta D} \neq 0$ & 100 & 1.595 & 24.1 &126\\ \hline
\end{tabular}
\end{table}

However, when both $\mathbf{R^{0}}$ and $\mathbf{D}$ are manipulated together ($\mathbf{\Delta D} \neq 0$), the impact becomes significantly more severe. The post-attack operation cost rises sharply to $1.595 \times 10^5~\$/h$, an increase of $24.1 \times 10^3~\$/h$, and load shedding escalates to $126$ MW. 

These results highlight that coordinated manipulation of solar generation and load exacerbates both economic and operational disruptions. Compared to isolated manipulation of $\mathbf{R^{0}}$, the coordinated attack intensifies grid strain, demonstrating the need for robust monitoring of both load and solar generation data.

\section{Conclusion}\label{sec:conclusion}
In this work, we propose a novel LR attack model that integrates the manipulation of distributed solar generation into the existing framework to highlight the grid's vulnerabilities resulting from cyber attacks targeting the increasing penetration of distributed solar generation. This model considers an assumption that generator measurements are no longer immune to tampering, a premise increasingly challenged by the growing integration of solar generation. We demonstrate that solar generation manipulation introduces significant vulnerabilities in the grid's economic operation, particularly when compared to scenarios that exclude solar generation.

Our analysis further reveals that the impact of manipulation varies with daily and seasonal patterns, with the grid being most vulnerable during periods of peak solar generation. These findings underscore the critical need for enhanced security mechanisms to safeguard modern power grids against emerging cyber threats.

For future work, we will enhance the proposed LR attack model by incorporating \textit{communication delays} and \textit{dynamic response analysis} to provide a more precise and realistic impact assessment of LR attacks. These enhancements will help address the evolving cyber security challenges in the smart grid landscape.

\vspace{0.2cm}
\section*{Acknowledgment}\label{sec:acknow}
This work was supported by the National Science Foundation under award numbers 2301938 and 2418359, and by Hitachi America. 

\bibliographystyle{ieeetr}
\bibliography{manuscript}
\end{document}